\documentstyle[12pt,epsfig]{ioplppt}
\begin{document}
\newcommand{\asin}{\sin^{-1}}
\newcommand{\erf}{\, \mathrm{erf}}
\jl{1}
\letter{Weight decay induced phase transitions in multilayer neural networks}
\author{M Ahr, M Biehl and E Schl\"osser \\ 
Institut  f\"ur Theoretische Physik \\
Julius-Maximilians-Universit\"at W\"urzburg, Am Hubland \\
D-97074 W\"urzburg, Germany}
\maketitle
\begin{abstract}
We investigate layered neural networks with differentiable activation 
function and student vectors without normalization constraint by means of 
equilibrium statistical physics. 
We consider the learning of perfectly realizable rules and find that the 
length of student vectors becomes infinite, unless a proper weight 
decay term is added to
the energy. Then,
the system undergoes a first order phase transition between states with 
very long student vectors and states where the lengths are comparable 
to those of the teacher vectors. Additionally in both configurations 
there is a phase transition between a specialized and an unspecialized phase. 
An anti-specialized phase with long student vectors exists in 
networks with a small number of hidden units. 
\end{abstract}
\medskip
\setlength{\unitlength}{1,1cm}
Statistical physics has been applied successfully to the investigation of
equilibrium states of neural networks. \cite{Hertz:Krogh:Palmer,Bishop} 
The by now standard analysis of 
off-line training from a fixed training set is based on the interpretation
of training as a stochastic process which leads to a well-defined thermal 
equilibrium. Investigations of perceptrons \cite{Seung:Sompolinsky:Tishby,
Watkin:Rau:Biehl,Boes:Kinzel:Opper} or 
committee machines \cite{Schwarze:Hertz,Schwarze,Opper,Schottky, Urbanczik}
have widely improved understanding of learning in neural networks. Meanwhile 
these studies are being extended to the more application relevant scenario 
of networks with continuous activation function and output. 
\cite{Biehl:Schloesser:Ahr,Ahr:Biehl:Urbanczik,Kang:Oh:Kwon:Park}

The soft-committee machine is a two-layered neural network which consists 
of a layer of $K$ hidden units, all of which are connected with the entire
$N$-dimensional input $\underline{\xi}$. The total output $\sigma$ ist 
proportional to the 
sum of outputs of all hidden units:
\begin{equation}
\textstyle{\sigma(\underline{\xi}) = \frac{1}{\sqrt{K}} \sum_{j = 1}^{K} g(x_{j}) \, \, \,
\mbox{where}\, \, \,  x_{j} = \frac{1}{\sqrt{N}} \underline{J}_{j} 
\cdot \underline{\xi}}
\end{equation}
where the weights of the $j$-th hidden unit are represented by the 
$N$-dimensional vector $\underline{J}_{j}$. 
We investigate learning of a perfectly matching rule parametrized
by a teacher network of the same architecture with output $\tau$ and
orthogonal vectors $\underline{B}_{j}$, which we assume to have the length
$\sqrt{N}$. 
The transfer function $g(x)$ is taken to be a sigmoidal function, e.g. the 
error function. 
Networks of this type have been studied in the limit of  high temperature 
\cite{Biehl:Schloesser:Ahr}, the annealed approximation \cite{Kang:Oh:Kwon:Park}, and by means of the replica
formalism \cite{Ahr:Biehl:Urbanczik}. All these studies imposed the simplifying condition that the order parameters
$Q_{ij} = \underline{J}_{i} \cdot \underline{J}_{j} / N$ are restricted to the
value 1 for $ i = j $, so the length of the student vectors is
fixed to that of the teacher vectors. This system shows a phase transition
between an unspecialized configuration, where the student-teacher overlaps 
$R_{ij} = \underline{J}_{i} \cdot \underline{B}_{j} / N $ are identical for 
all $i, j$ and a specialized configuration where $R_{ii} \not = R_{ij}$
for $i \not = j$. However, constraining the student lengths implies significant
{\it a priori} knowledge of the rule which is not available in practical 
applications. So, in this paper we want to obtain first results for
soft committee machines  which determine student lengths in the course 
of learning.   

Learning is guided by the minimization of the training error
\begin{equation}
\textstyle{\epsilon_{t} = \frac{1}{P} \sum_{\mu = 1}^{P} \frac{1}{2}
\left( \sigma(\underline{\xi}_{\mu}) - \tau(\underline{\xi}_{\mu}) \right)^{2}}
\end{equation}
where $P$ is the number of examples used for training.
After training, the success of learning can be quantified by an average 
of the quadratic error measure over the distribution of possible inputs,
the so-called generalization error:
\begin{equation}
\textstyle{\epsilon_{g} = \frac{1}{2}\left<  \left(  \sigma(\underline{\xi}) - 
\tau(\underline{\xi}) \right)^{2} \right>_{\underline{\xi}}}
\end{equation}
Following the standard statistical physics approach, we consider a Gibbs
ensemble, which is characterized by the partition function
$
Z = \int d\mu(\{J_{i}\}) \, \exp(- \beta H (\{\underline{J}_{i}\}))
$ 
with a formal temperature $1/\beta$ which controls the thermal average
of energy in the equilibrium. The extensive energy $H$ is a
function of the training error, the standard choice being  
$H = P \epsilon_{t}$.  Typical equilibrium properties are calculated from the associated 
quenched free energy $-(1/\beta) \left< \ln Z \right> =: f N$ where the average 
is performed over
the random set of training examples. The evaluation
of $\left< \ln Z \right>$ in general requires the rather involved replica 
formalism. To obtain first results we consider the simplifying 
{\it high-temperature limit} $\beta \rightarrow 0$ \cite{Seung:Sompolinsky:Tishby,Watkin:Rau:Biehl}.
The calculation of equilibrium states is guided by minimization of
$
\beta f = \tilde{\alpha} K \epsilon_{g} - s
$. 
Here $\widetilde{\alpha} = \beta P / (N K)$ ist the rescaled number of examples, 
which we assume to be ${\mathcal O} (1)$ and $s$ the entropy per degree of
freedom with order parameters held fixed. The latter is given by
\begin{equation}
s = 1/2 \ln \det \mat{C} + \mbox{irrelevant const.}
\label{entropy}
\end{equation}
where $\mat{C}$ is the $2 K \times 2 K$-matrix of all cross- and self-overlaps
of student and teacher vectors. Equation \ref{entropy} is of quite general validity
and can be derived by means of a saddle point integration from the definition of the 
entropy. In \cite{Ahr:Biehl:Urbanczik} a simpler derivation is presented.

Here we assume the components of all examples to be independent random
numbers with mean zero and unit variance. 
Then, in the thermodynamic limit $N \rightarrow \infty$ the generalization
error can be calculated analytically, if we choose the activation function
$g(x) = \erf(x/\sqrt{2})$ \cite{Biehl:Schwarze,Saad:Solla} which is very similar to the more popular
hyperbolic tangent, so the basic features of the model should not be altered:
\begin{equation}
\textstyle{\epsilon_{g} = \frac{1}{6} + \frac{1}{K \pi} \sum_{i, k = 1}^{K} \left[
\asin \left( \frac{Q_{ik}}{\sqrt{(1 + Q_{ii})(1 + Q_{kk})}} \right)
- 2 \asin \left( \frac{R_{ik}}{\sqrt{2 ( 1 + Q_{ii})}} \right) \right]}
\label{generalization}
\end{equation}

In the following, we will first investigate the simplest case $K = 1$, i.e.
 a network consisting of one single unit to show the basic principles. 
Then we will study networks with arbitrary $K$ and finally  
investigate the limit $K \rightarrow \infty$ of very large networks.

In the $K=1$ case equations \ref{entropy} and
 \ref{generalization} read:
\begin{eqnarray}
\epsilon_{g} &=& \textstyle{\frac{1}{6} +
\frac{1}{\pi} \asin \left( \frac{Q}{1 + Q} \right) -
\frac{2}{\pi} \asin \left( \frac{R}{\sqrt{2(1 + Q)}} \right) }\\
s &=& \textstyle{\frac{1}{2} \ln \left( Q - R^{2} \right)}
\end{eqnarray}
Trying to minimize $\widetilde{\alpha} \epsilon_{g} - s$, we find that 
$\epsilon_{g}$ remains of order 1 for arbitrary $R$, $Q$ while $s$ becomes
infinite for $Q \rightarrow \infty$, yielding $f \rightarrow - \infty$.
This means that in thermal equilibrium the length of the student vector
increases to infinity, while its overlap with the teacher becomes irrelevant.
Of course, this is not the desired result of training. 
The method of choice to avoid this behavior, is to ``punish'' configurations
with large $Q$ with an additional energy called ``weight 
decay''. This is a method of {\it regularization} which is widely 
used in practice in order to improve the generalization ability of feedforward 
neural networks \cite{Hertz:Krogh:Palmer}.  So we introduce 
$H = P \epsilon_{t} + \lambda N Q$ \cite{Boes:Opper,Sollich,Dunmur:Wallace,
Krogh:Hertz,Saad:Rattray,Barber:Sollich} and obtain $\beta f = \widetilde{\alpha}
\epsilon_{g} + \widetilde{\lambda} Q - s$ with $\widetilde{\lambda} = \beta \lambda$
which has to be minimized w.r.t. $R$ and $Q$. 
\begin{figure}
\begin{center}
\begin{picture}(14,4.5)(0,0)
\put(0,-1.5){\resizebox{7cm}{8cm}{\includegraphics{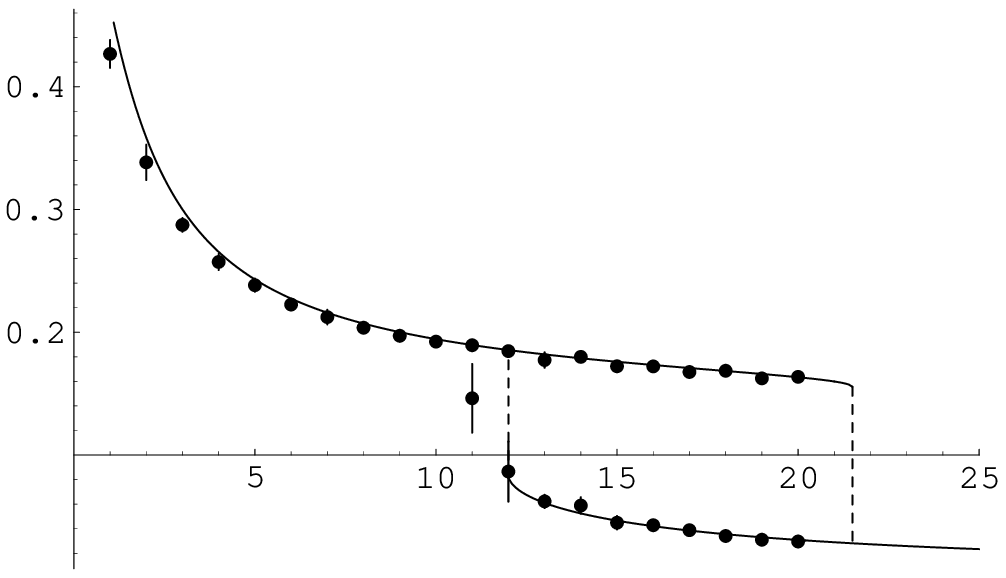}}}
\put(3.5,1.5){\resizebox{3cm}{4cm}{\includegraphics{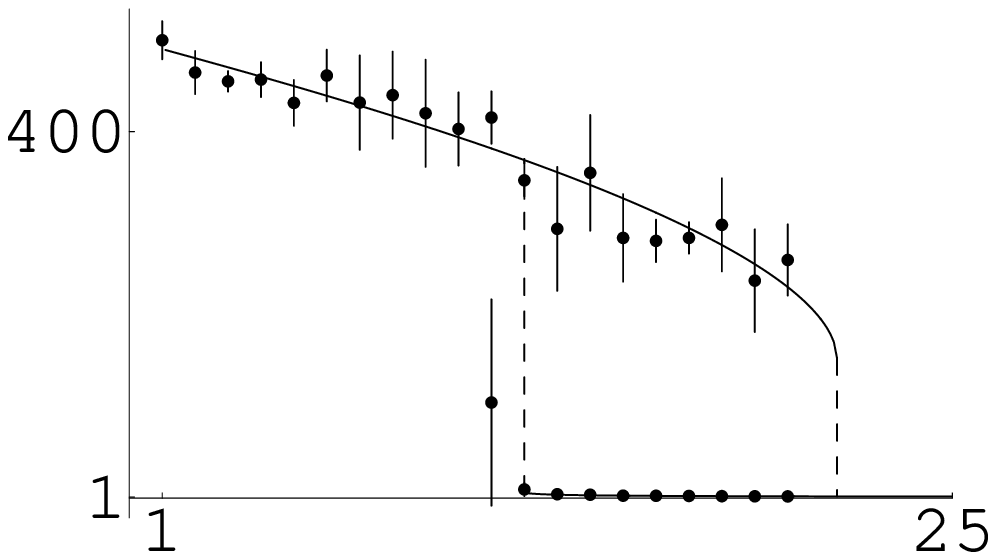}}} 
\put(7,-1.5){\resizebox{7cm}{8cm}{\includegraphics{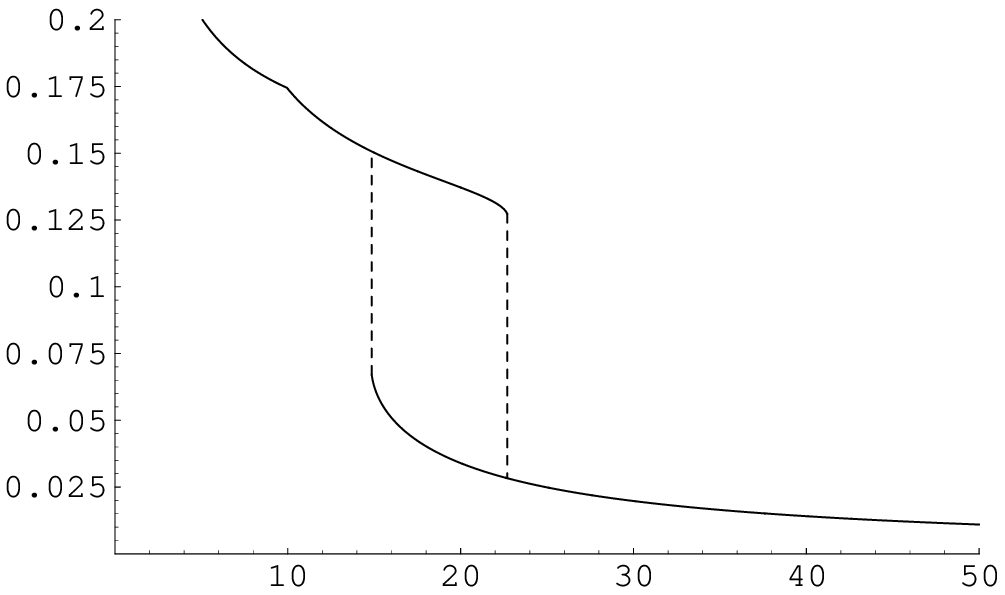}}}
\put(10.5,1.0){\resizebox{3cm}{4cm}{\includegraphics{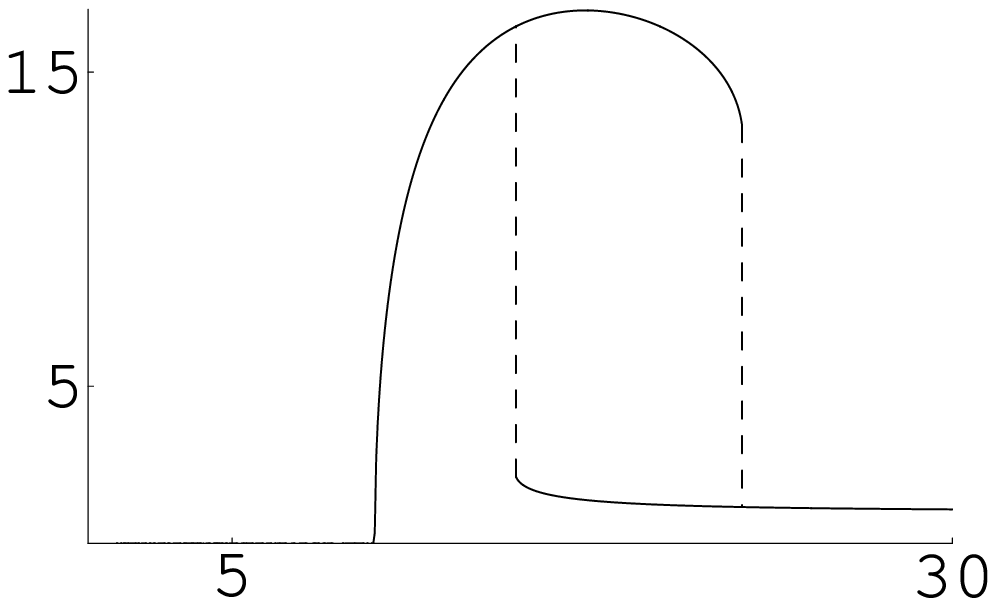}}}
\put(6.5,0){$\widetilde{\alpha}$}
\put(13.5,0){$\widetilde{\alpha}$}
\put(0,4.5){$\epsilon_{g}$}
\put(7,4.5){$\epsilon_{g}$}
\put(6,2.0){$\widetilde{\alpha}$}
\put(3.5,4.5){$Q$}
\put(13,1.5){$\widetilde{\alpha}$}
\put(10.5,4){$\Delta$}
\end{picture}
\end{center}
\caption{ left panel:  $\epsilon_{g}(\widetilde{\alpha})$ and $Q(\widetilde{\alpha})$
(inset)
as obtained analytically (solid line) and results of Monte-Carlo simulations 
(dots) for $K = 1$, $\widetilde{\lambda} = 0.001$ and $\beta = 0.2$. 
(system size $N = 100$, averages over 5 runs with 10000 M.C. steps each, 
5000 of which were used for equilibration, 5000 for sampling measurements) 
We get two locally stable states 
with different student lengths for some $\widetilde{\alpha}$, depending on the 
starting value of the student vector. We have used the same strategy as in \cite{Schwarze:Hertz:2} to obtain the hysteresis behaviour. 
right panel: $\epsilon_{g}(\widetilde{\alpha})$ and $\Delta(\widetilde{\alpha})$
(inset) for $K = 2$ and $\widetilde{\lambda} = 0.001$.}
\label{kaeins}
\end{figure}
In Figure \ref{kaeins} we show $\epsilon_{g}$ as a function of the rescaled
number of examples, $\widetilde{\alpha}$ for $\widetilde{\lambda} = 0.001$. For small 
$\widetilde{\alpha}$ the network is in a state with large $Q$ 
(and $\epsilon_{g}$). For  $\widetilde{\alpha} \geq 12$ a second state with 
small $Q$ and small $\epsilon_{g}$ exists, which becomes globally stable 
at $\widetilde{\alpha} \approx 15$. At $\widetilde{\alpha} \approx 21.6$ the state with large
 $Q$
becomes even locally unstable. We remark that this phase transition
is solely due to the differentiable nature of the activation function, which
causes the energy to depend on the length of the student vector, and
does not occur in the simple perceptron. It was also not found for the simpler 
linear unit with $g(x) = x$ \cite{Dunmur:Wallace}, where the training error
is more sensitive to a mismatched $Q$ than in the case of a bounded, 
saturating transfer function. The phase transition disappears for $\tilde{\lambda} \geq 0.006$.  

We have performed continuous Monte-Carlo simulations of a Metropolis-like
learning process of the single unit. The results shown in Figure \ref{kaeins} 
confirm our theoretical results. 

\begin{figure}
\begin{center}
\begin{picture}(14,4.5)(0,0)
\put(0,-1.5){\resizebox{7cm}{8cm}{\includegraphics{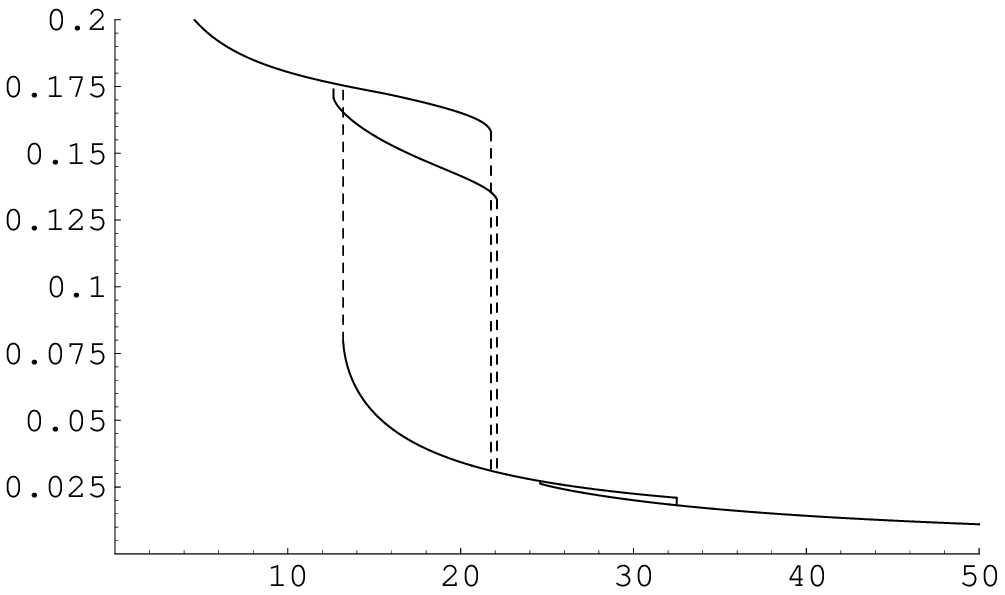}}}
\put(3.5,0.5){\resizebox{3cm}{4.5cm}{\includegraphics{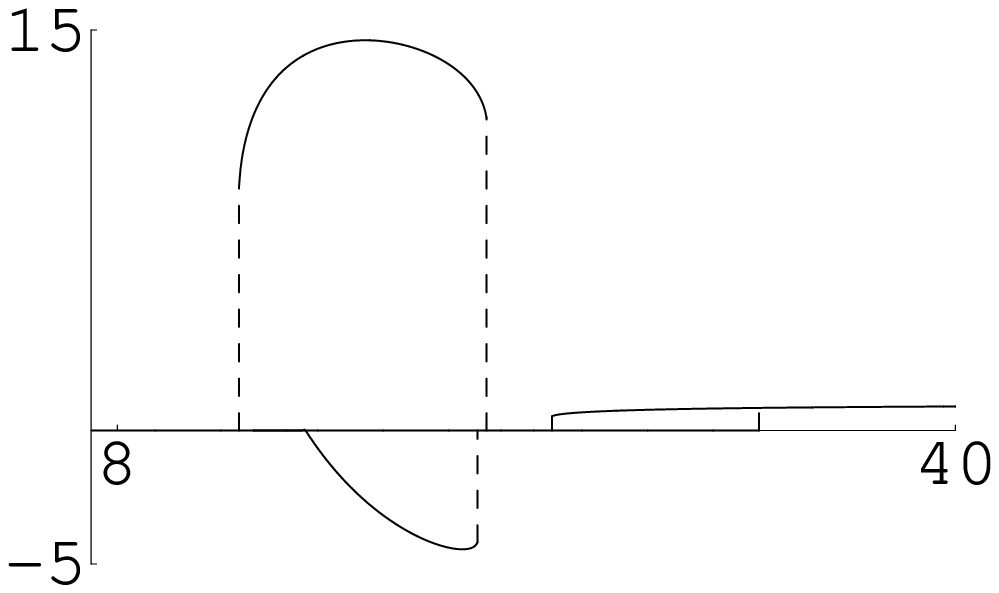}}} 
\put(7,-1.5){\resizebox{7cm}{8cm}{\includegraphics{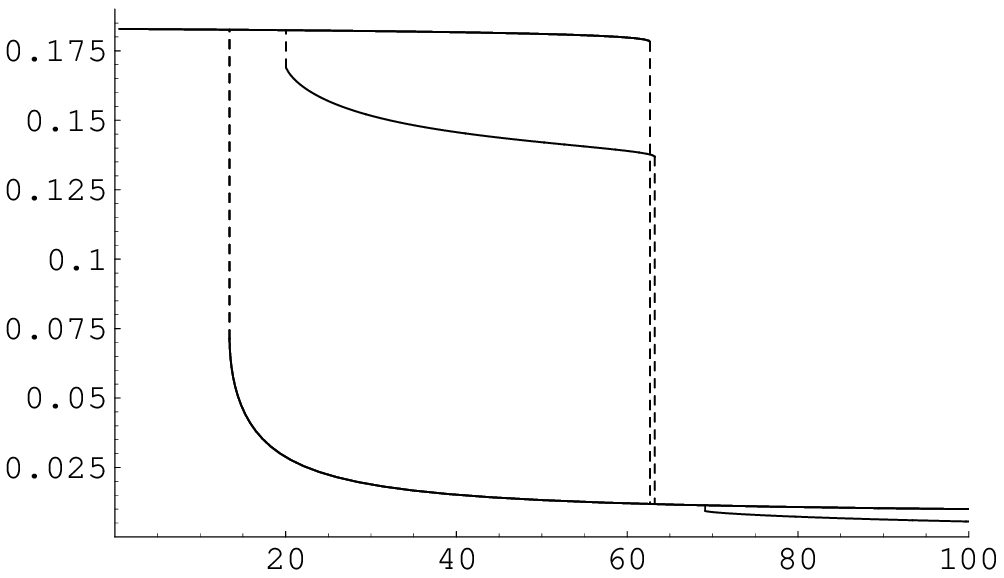}}}
\put(6.5,0){$\widetilde{\alpha}$}
\put(13.5,0){$\widetilde{\alpha}$}
\put(0,4.5){$\epsilon_{g}$}
\put(7,4.5){$\epsilon_{g}$}
\put(6,1.5){$\widetilde{\alpha}$}
\put(3.5,4){$\Delta$}
\end{picture}
\end{center}
\caption{left: $\epsilon_{g}(\widetilde{\alpha})$ for $K = 3$, 
$\widetilde{\lambda} = 0.001$ and $\Delta(\widetilde{\alpha})$ (inset). Different
starting values were used in numerical minimization to calculate as many local
minima of the free energy as possible. 
right: $K = \infty$, $\widetilde{\lambda} = 0.0001$. 
All local minima of the free energy have been calculated from the saddle point
equations.} 
\label{kazwei}
\end{figure}

In order to extend our analysis to networks with $K \geq 2$ we assume the 
network configuration to be {\it site-symmetric} with respect to the hidden units
so the order parameters fulfill the conditions 
$R_{ij} = R \delta_{ij} + S (1 - \delta_{ij})$ and 
$Q_{ij} = Q \delta_{ij} + C (1 - \delta_{ij})$. This assumption reflects the symmetry 
of the rule yet allows for specialization of the student, as student overlaps 
with teacher vectors can yield different values for $i = j$ and $i \not = j$.
Now generalization error and entropy read:
\begin{equation}
\fl \textstyle{ \epsilon_{g} = \frac{1}{6} + \frac{1}{\pi} \asin \left( \frac{Q}{1 + Q}
\right) + \frac{K - 1}{\pi} \left[ \asin \left( \frac{C}{1 + Q} \right)
- 2 \asin \left( \frac{S}{\sqrt{2(1 + Q)}} \right) \right] - \frac{2}{\pi} 
\asin \left( \frac{R}{\sqrt{2(1 + Q)}} \right)}
\end{equation}
\begin{equation}
\fl \textstyle{s = \frac{1}{2} \ln \left[ (K - 1) C + Q - (R + (K - 1) S)^{2}
\right] + \frac{K - 1}{2} \ln \left[ Q - C - (R - S)^2 \right]}
\end{equation}
The weight decay term introduced for the single unit generalizes naturally
to $\lambda \sum_{i = 1}^{K} Q_{ii}$, so the free energy becomes 
$\beta f = \widetilde{\alpha} K \epsilon_{g} + 
\widetilde{\lambda} K Q - s$ in a site symmetric state. Numerical minimization  
leads to the results shown in Figure
 \ref{kaeins} and \ref{kazwei} for $K = 2$ and $K = 3$. 
In addition to the first order phase transition already observed at the 
single unit, which connects states with different lengths of student vectors, 
we observe transitions between phases which are characterized by
the parameter $\Delta := R - S$ indicating specialization features.
As both transitions are due to independent 
mechanisms, namely on the one hand a change of student vector {\it lengths} and
on the other hand  an alteration of their {\it directions}, specialized 
($\Delta > 0$)
and unspecialized ($\Delta = 0$) phases can exist both in the large-$Q$ 
configuration and
in the small-$Q$ regime. Indeed for $K \geq 3$ first order transitions
between specialized and unspecialized phases can be observed in both 
configurations. Additionally, there is a second order phase transition
between the unspecialized large $Q$ phase and an anti-specialized phase 
($\Delta < 0$) with large $Q$ 
at $\widetilde{\alpha} \approx 15$.
The $K = 2$ system shows a second order transition
in the large-$Q$ regime, while an unspecialized configuration with small
$Q$ cannot be observed. This difference in behavior results from the higher 
degree of symmetry in the $K = 2$ system, where the free energy is invariant
under exchange of $R$ and $S$. Consequently  there is no physical difference 
between specialized and anti-specialized configurations in the $K = 2$ system.

To study the behaviour of very large networks ($K \rightarrow \infty$) 
scaling assumptions of order parameters have to be made. 
Supposing $C$ to be ${\mathcal O}(1)$, the output of the student will be 
${\mathcal O}(\sqrt{K})$ and thus on a different scale as the teacher 
output. So we assume the hidden unit overlaps to be ${\mathcal O}(1/K)$, 
writing $C = \hat{C}/(K - 1)$ and further introduce $S = \hat{S}/K$, while
$\Delta$ and $Q$ remain ${\mathcal O}(1)$. Inserting this and performing 
$\lim_{K \rightarrow \infty} \beta f /K$ we find that the condition
$\partial f/\partial S = 0$ can be fulfilled only if $Q + \hat{C} - (\Delta 
+ \hat{S})^{2}$ is assumed to be ${\mathcal O}(1/K)$.
So we substitute $ \widehat{C} = \widetilde{C}/K +
 (\Delta + \widehat{S})^{2} - Q$ before
performing the limit $K \rightarrow \infty$. 
The corresponding generalization error is shown in Figure \ref{kazwei}
as a function of $\widetilde{\alpha}$. For small $\widetilde{\alpha}$, 
the network is in an unspecialized phase with large $Q$. At $\widetilde{\alpha} \approx
13$ a locally stable, unspecialized configuration with small $Q$ appears,
which is globally stable between  $\widetilde{\alpha} \approx 22$ and 
$\widetilde{\alpha} \approx  88$, where the specialized small $Q$ configuration becomes
globally stable. However, the unspecialized configuration remains locally 
stable. Additionally, at $\widetilde{\alpha} \approx 20$ the specialized large $Q$
phase appears, the free energy of which is smaller than that of the 
unspecialized large $Q$ phase for $\widetilde{\alpha} > 22.5$.
Anti-specialized configurations do not exist in the limit $K \rightarrow
\infty$. We expect them to be a characteristic feature of 
systems with small $K \geq 3$. 

In summary, we have shown by means of statistical physics that learning 
an unknown rule without a priori knowledge in the form of normalized 
student vectors leads to a much more complicated behaviour than learning
with normalized students. The number of phases in which the system can
exist increases. Further, student lengths tend to infinity unless the network 
weights are regularized  by means of a proper weight decay.

Further investigations will extend research to finite temperatures
by applying the replica formalism and study the relevance of our results
for practical training processes.

\ack

We thank W Kinzel and A Freking for stimulating discussions and a critical
reading of the manuscript.

\end{document}